\documentclass[preprint,showpacs]{revtex4}
\usepackage{times,xspace}
\usepackage{amsbsy,amssymb,amsmath,bm}
\usepackage{fancyheadings}
\usepackage{graphicx,color,epsfig,rotate}

\def\bbbc{{\mathchoice {\setbox0=\hbox{$\displaystyle\rm C$}\hbox{\hbox
to0pt{\kern0.4\wd0\vrule height0.9\ht0\hss}\box0}}
{\setbox0=\hbox{$\textstyle\rm C$}\hbox{\hbox
to0pt{\kern0.4\wd0\vrule height0.9\ht0\hss}\box0}}
{\setbox0=\hbox{$\scriptstyle\rm C$}\hbox{\hbox
to0pt{\kern0.4\wd0\vrule height0.9\ht0\hss}\box0}}
{\setbox0=\hbox{$\scriptscriptstyle\rm C$}\hbox{\hbox
to0pt{\kern0.4\wd0\vrule height0.9\ht0\hss}\box0}}}}

\newcommand{\ignore}[1]{}
\newcommand{\mComment}[1]{}
\newcommand{\gComment}[1]{}
\newcommand{\jComment}[1]{}
\newcommand{\rComment}[1]{}
\newcommand{\lComment}[1]{}
\renewcommand{\mComment}[1]{\textcolor{blue}{Manny: #1}}
\renewcommand{\gComment}[1]{\textcolor{red}{Gerardo: #1}}
\renewcommand{\jComment}[1]{\textcolor{green}{Jim: #1}}
\renewcommand{\rComment}[1]{\textcolor{magenta}{Ray: #1}}
\renewcommand{\lComment}[1]{\textcolor{purple}{Rolando: #1}}
\input{tcilatex}

\begin{document}

\title{Magnetic Field-Induced Condensation of Triplons in Han Purple Pigment
BaCuSi$_2$O$_6$}
\author{M. Jaime$^{1*}$, V. F. Correa$^1$, N. Harrison$^1$, C. D. Batista$^2$%
, N. Kawashima$^3$, Y. Kazuma$^3$, G. A. Jorge$^{1,4}$, R. Stern$^5$, I.
Heinmaa$^5$, S. A. Zvyagin$^6$, Y. Sasago$^{7\dagger}$, K. Uchinokura$%
^{7\dagger\dagger}$}
\affiliation{$^1$ National High Magnetic Field Laboratory, MS-E536, Los Alamos, NM 87545,
USA.}
\affiliation{$^2$ Theoretical Division, Los Alamos, NM 87545, USA.}
\affiliation{$^3$ Department of Physics, Tokyo Metropolitan University, Tokyo 192-0397,
Japan.}
\affiliation{$^4$ Departmento de F\'{\i}sica, Universidad de Buenos Aires, Argentina.}
\affiliation{$^5$ National Institute of Chemical Physics and Biophysics, 12618 Tallinn,
Estonia.}
\affiliation{$^6$ National Magnetic Field Laboratory, Tallahassee, FL 32310, USA.}
\affiliation{$^7$ Department of Applied Physics, The University of Tokyo, Tokyo,
113-8656, Japan.}

\begin{abstract}
Besides being an ancient pigment, BaCuSi$_2$O$_6$ is a quasi-2D magnetic
insulator with a gapped spin dimer ground state. The application of strong
magnetic fields closes this gap creating a gas of bosonic spin triplet
excitations called triplons. The topology of the spin lattice makes BaCuSi$_2
$O$_6$ an ideal candidate for studying the Bose-Einstein condensation
of triplons as a function of the external magnetic field, which acts as a
chemical potential. In agreement with quantum Monte Carlo numerical
simulation, we observe a distinct lambda-anomaly in the specific heat together
with a maximum in the magnetic susceptibility upon cooling down to liquid
Helium temperatures.
\end{abstract}

\maketitle

%\address{$^1$ National High Magnetic Field Laboratory, MS-E536, Los Alamos, NM 87545, USA. \\
%$^2$ CNLS-MS B258, Los Alamos, NM 87545, USA.\\
%$^3$ Department of Physics, Tokyo Metropolitan University, Tokyo 192-0397, Japan. \\
%$^4$ Departmento de F\'{\i}sica, Universidad de Buenos Aires, Argentina. \\ 
%$^5$ National Institute of Chemical Physics and Biophysics, 12618 Tallinn, Estonia. \\
%$^6$ National Magnetic Field Laboratory, Tallahassee, FL 32310, USA. \\
%$^7$ Department of Applied Physics, The University of Tokyo, Tokyo, 113-8656, Japan.}

About 2000 years ago, early Chinese chemists synthesized barium copper
silicates for the first time, and used them as pigments for pottery and
trading as well as for large Empire projects such as the Terracotta Warriors 
\cite{FitzHugh,Berke,Zuo}, preceding even the invention of paper and the
compass. BaCuSi$_2$O$_6$, also known as Han Purple \cite{FitzHugh}, is then
possibly the first man-made compound containing a metallic bond. In its
layered crystallographic structure, pairs of Cu$^{2+}$ ions form dimers
arranged in a square lattice, with Cu-Cu bonds projected normal to the
planes \cite{Finger}. Neighboring Cu bilayers are weakly coupled, i.e. the
magnetic system is quasi-two dimensional \cite{Sasago}. Applied magnetic
fields in excess of $23.5 \mathrm{T}$ suppress the spin singlet ground state
in BaCuSi$_2$O$_6$, giving rise to a gas of quantum spin-triplet excitations.

At low temperatures, an ideal gas of bosons undergoes a phase transition
into a condensate with macroscopic occupation of the single-particle ground
state. A great deal of interest in this phenomenon was triggered by the
discovery of the anomalous behavior of liquid helium. Upon cooling, liquid $%
^{4}$He exhibits a $\lambda $-transition in the specific heat at 2.17~K that
signals the onset of a zero-viscosity superfluid state \cite{London}.
Superfluidity arises from the macroscopic fraction of He atoms that occupy
the single particle ground state. Within the last decade, Bose-Einstein
condensation (BEC) was also realized for dilute clouds of atoms at
temperatures lower than ten millionths of a degree Kelvin \cite{Pethick}.
The possibility of producing BEC with quantum spin magnets\cite%
{matsubara-afleck} has stimulated considerable experimental effort to find
the candidate materials\cite{Rice}.

Spin-dimer systems such as SrCu$_{2}$(BO$_{3}$)$_{2}$ \cite{Kageyama} and
TlCuCl$_{3}$ \cite{Oosawa99} have recently become of interest owing to the
ability of strong magnetic fields to generate a gas of $S^{z}=1$
spin-triplet states moving in a non-magnetic background. These \textit{%
triplons} \cite{Schmidt}can be regarded as bosonic particles with hard core
repulsions that carry a magnetic moment, but no mass or charge. The external
field plays the role of a chemical potential in controlling the number of
particles. The Ising component of the inter-dimer exchange interaction
generates an effective repulsion between neighboring bosons.

Upon cooling, the gas of triplons can either crystallize or condense in a
liquid state, depending on the balance between the kinetic energy and the
repulsive interactions \cite{Rice}. In particular, if the kinetic energy
dominates and the number of triplons (or total magnetization) is a conserved
quantity, the system undergoes BEC corresponding to the coherent
superposition of $S^z~=~1$ spin-triplet and S~=~0 singlet states on each and
every dimer. The BEC of triplons was proposed to occur in TlCuCl$_3$ \cite%
{Oosawa99,Nikuni,Ruegg} and KCuCl$_3$ \cite{Cavadini,Oosawa02}. However,
recent data suggest that the anisotropic crystal structure of these
spin-ladder systems breaks the U(1) rotational invariance that is required
to have gapless Goldstone modes \cite{Glazkov}. Hence, the Goldstone modes
acquire a finite gap or mass. In this report we introduce a new system whose
bi-layer structure provides a realization of a quasi-two dimensional gas of
triplons.

The range of fields and temperatures required to obtain a BEC of triplets in
BaCuSi$_2$O$_6$ creates a unique situation from the experimental standpoint:
the triplons dominate the thermodynamics in a very simple and predictable
fashion. The simplicity of BaCuSi$_2$O$_6$ is best described by a spin
Hamiltonian that includes only the nearest-neighbor Heisenberg
antiferromagnetic exchange couplings. The Cu$^{2+}$ ions and the SiO$_4$
tetrahedra in BaCuSi$_2$O$_6$ are arranged in layers parallel to the $(001)$
crystallographic plane, as shown in Fig. 1(a). Within each Si-O-Cu
layer, the Cu$^{2+}$ ions form a square lattice bi-layer of $S = 1/2$ spins 
\cite{Finger}. The relevant Heisenberg exchange couplings within the
bi-layers are the intra-dimer interaction, $J$, and an inter-dimer
nearest-neighbor interaction $J^{\prime}$ (Fig.~1(a)). The different
bi-layers are antiferromagnetically coupled via the effective exchange
constant $J^{\prime\prime}$. The resulting spin Hamiltonian is:

\begin{eqnarray}
H_{s} &=&J\sum_{\mathbf{i}}\mathbf{S}_{\mathbf{i},1}\cdot \mathbf{S}_{%
\mathbf{i},2}+J^{\prime }\sum_{\mathbf{i,\alpha ,\beta }}\mathbf{S}_{\mathbf{%
i},2}\cdot \mathbf{S}_{\mathbf{i}+{\hat{e}}_{\alpha },2}  \notag \\
&+&J^{\prime \prime }\sum_{\mathbf{i}}\mathbf{S}_{\mathbf{i},1}\cdot \mathbf{%
S}_{\mathbf{i}+{\hat{z}},2}-g_{\parallel }\mu _{B}H\sum_{\mathbf{i,\beta }}S_{\mathbf{%
i},\beta }^{z}
\end{eqnarray}%

where $\alpha =\{1,2\}$ is the direction index, $\beta =\{1,2\}$ is the
layer index, ${\hat{e}}_{1}={\hat{x}}$, ${\hat{e}}_{2}={\hat{y}}$ and ${\hat{%
z}}$ are unit vectors along the crystallographic axes, and $g_{\parallel
}=2.306\pm 0.008$ is the component of the gyromagnetic tensor along the $c$%
-axis, which is the direction of the applied magnetic field $H$. The value
of $J=4.5\mathrm{meV}$ is known from inelastic neutron scattering studies. 
\cite{Sasago}, which also show that $J^{\prime \prime }\ll J^{\prime }\ll J$.
Precise values of $J^{\prime }$ and $J^{\prime \prime }$ are determined
from the experiments described below.

\begin{figure}[htb]
\vspace{-2.2cm} 
\includegraphics[angle=270,width=6.5cm,scale=1]{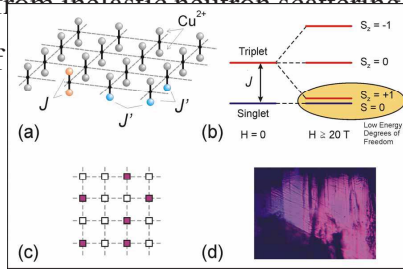}
\vspace{-0.8cm}
\caption{(color online) a) Cu$^{2+}$-dimer plane in BaCuSi$_2$O$_6$ in a
square lattice arrangement. $J$ and $J^{\prime}$ are the intra-dimer and
inter-dimer exchange constants. b) Evolution of the isolated dimer spin
singlet and triplet with an applied magnetic field H. c) Partially occupied
hard core boson lattice. Occupation (chemical potential) is given by the
external magnetic field. d) Transmission-light picture of a BaCuSi2O6 single
crystal.}
\label{fig1}
\end{figure}

In the limit $J^{\prime}=J^{\prime\prime}=0$, the ground state of $H_s$ is a
product of local singlet-dimer states: $|\Psi_0 \rangle = \otimes_{\mathbf{i}%
} |\phi_{\mathbf{i}}^s \rangle$ where $|\phi_{\mathbf{i}}^s \rangle$ is the
singlet state between two spins on a dimer $\mathbf{i}$. $J = 4.5~\mathrm{meV}$
is the energy gap for a spin triplet excitation in a single isolated dimer 
\cite{Sasago}. This gap decreases linearly with an applied magnetic field as
shown in Figure 1(b). Since $J^{\prime \prime }, J^{\prime }\ll J$, 
the spectrum is dominated by the two low energy states of the isolated dimer 
(the $S~= 0$ singlet and the $S^z = 1$ triplet) once $g_{\parallel} \mu_B H$ 
becomes of the order of $J$. By projecting $H$ onto this low energy subspace, 
we derive an effective Hamiltonian, $H_{\text{\emph{eff}}}$, 
in which each dimer is represented by one effective site with two possible states.
In this way, we are neglecting the effect of the high energy triplet states
and, therefore, the highest order terms of $H_{\text{\emph{eff}}}$ are 
linear in $J^{\prime}$ and $J^{\prime\prime}=0$ (first order perturbation
theory). A similar procedure was used before by
Mila \cite{Mila} to describe spin ladders in a magnetic field. We 
associate the two low energy states with the two possible states of a hard
core boson on a lattice \cite{Sasago} (see Fig.~1(c)). The empty site
corresponds to the singlet state $|\phi_{\mathbf{i}}^s \rangle$, while the
site occupied by the hard core boson represents the $S^z = 1$ triplet state:
$|\phi_{\mathbf{i}}^s \rangle \rightarrow |0 \rangle_{\mathbf{i}}$,   
$|\phi_{\mathbf{i}}^t , S^{z}=1 \rangle \rightarrow b^{\dagger}_{\mathbf{i}}
|0\rangle_{\mathbf{i}}$,where $|0\rangle_{\mathbf{i}}$ represents the empty 
state at the site $\mathbf{i}$. In this language, the effective low energy 
Hamiltonian is:

\begin{eqnarray}
H_{\text{\emph{eff}}} &=& t \sum_{\mathbf{i, \alpha}} (b^{\dagger}_{\mathbf{i}+{\hat e%
}_{\alpha}}b^{\;}_{\mathbf{i}}+ b^{\dagger}_{\mathbf{i}} b^{\;}_{\mathbf{i}+{\hat e}%
_{\alpha}}) + t^{\prime}\sum_{\mathbf{i}} (b^{\dagger}_{\mathbf{i}+{\hat z}}b^{\;}_{%
\mathbf{i}}+ b^{\dagger}_{\mathbf{i}} b^{\;}_{\mathbf{i}+{\hat z}})  \notag \\
&+& V \sum_{\mathbf{i,\alpha }} n_{\mathbf{i}} n_{\mathbf{i}+{\hat e}_{\alpha}} +
V^{\prime}\sum_{\mathbf{i}} n_{\mathbf{i}} n_{\mathbf{i}+{\hat z}} + \mu \sum_{%
\mathbf{i}} n_{\mathbf{i}}  \label{Heff}
\end{eqnarray}

where the chemical potential is $\mu = J - g_{\parallel} \mu_B H$, $t = V =
J^{\prime}/2$, and $t^{\prime}= V^{\prime}= J^{\prime\prime}/4$. Eq.~\ref%
{Heff} describes a gas of hard core bosons with nearest-neighbor hopping and
repulsive interactions. For $H$ less than a critical field $H_{c1}$ ($%
g_{\parallel}\mu_{B}H_{c1} = J-2J^{\prime}-J^{\prime\prime}/2$), the ground
state of $H_{\text{\emph{eff}}}$ is completely empty, i.e. all dimers are
spin singlets, because the chemical potential, $\mu$, is large and positive.
A finite concentration of particles $\rho = m$ ($m$ is the magnetization per
site) emerges in the ground state only when $H$ exceeds $H_{c1}$. Because
the condition $V \leq 2t$ is not fulfilled, the triplets or hard core bosons
never crystallize for any concentration $\rho < 1$.

\begin{figure}[htb]
\vspace{-1.2cm} 
\includegraphics[angle=0,width=6.5cm,scale=1]{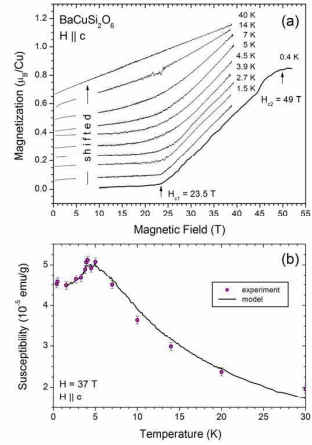}
\vspace{0.3cm}
\caption{(color online) a) Magnetization vs field at different temperatures
from $0.4 \mathrm{K}$ to $40~\mathrm{K}$, measured in both a $65~\mathrm{T}$
short-pulse, and a $45~\mathrm{T}$ mid-pulse, capacitor driven pulsed
magnets. b): Magnetic susceptibility vs temperature from measurements in A,
and from our model using a quantum Monte Carlo algorithm.}
\label{fig2}
\end{figure}

The single crystal samples used in this study were prepared using the
floating-zone technique. BaCO$_{3}$, CuO, and SiO$_{2}$ were used as
starting materials and the powders of stoichiometric compositions were mixed
and sintered at $700^{\circ }\mathrm{C}$ for $15\mathrm{h}$. The sintered
powder was reground and molded into a rod under a pressure of $400\mathrm{kg}/%
\mathrm{cm}^{2}$ for about $10\mathrm{min}$. The rod is sintered between $%
1000$ and $1100^{\circ }\mathrm{C}$ for about 100 h. The sintered rod is
then used for floating zone method \cite{Manabe}. The floating zone growth was
done twice. In the first run the speed of the growth was set to ~$50\mathrm{%
mm}/\mathrm{h}$, and in the second run to $\sim 0.5~\mathrm{mm}/\mathrm{h}$,
both in O$_{2}$ flow of $\sim 200~\mathrm{cc}/\mathrm{min}$. The purple
color and texture of our single crystal sample is appreciated in a picture
taken in transmission (backlight) mode (Fig.~1(d)). Figure~2%
a shows the magnetization versus field data obtained in capacitor-driven
pulsed magnetic fields \cite{Boebinger}. The magnetization increases
continuously between the critical field $H_{c1}=23.5~\mathrm{T}$ and the
saturation field $H_{c2}=49~\mathrm{T}$ at $T$ = 0.4~K. The nearly linear
slope of the magnetization contrasts sharply the staircase of plateaux
observed in SrCu$_{2}$(BO$_{3}$)$_{2}$, where the gas of triplets is known
to crystallize in a structure that is commensurate with the underlying
crystallographic lattice \cite{Rice,Kageyama}. Pulsed magnetic field
experiments on BaCuSi$_{2}$O$_{6}$ were repeated at different temperatures
up to 40~K and cross-calibrated against low-field SQUID magnetometer data 
\cite{Sasago}. The structure observed around $\simeq 48~\mathrm{T}$ at low
temperatures, which we could not reproduce in later experiments, is likely
an artifact caused by mechanical vibration of the probe. A plot
of the slope $\chi = dM/dH$ of $M$ versus $H$ (magnetic susceptibility)
at $H=37~\mathrm{T}$ is shown in figure~2(b) together with the
results of a Monte Carlo simulation of $H_{\text{\emph{eff}}}$ in a finite
lattice of size $L^{3}$, with $L=12$, performed using the directed-loop
algorithm \cite{Syljuasen,Harada} with parameters: $J=4.45~\mathrm{meV}$, $%
J^{\prime }=0.58~\mathrm{meV}$ and $J^{\prime \prime }=0.2J^{\prime }=0.116~%
\mathrm{meV}$. The agreement between experimental data and model
calculations is remarkable. Notably, the rapid drop of the magnetic
susceptibility below 4~K indicates the onset of an ordered low temperature
state, namely the triplon condensate.

\begin{figure}[htb]
\vspace{-2.2cm} 
\includegraphics[angle=270,width=6.5cm,scale=1]{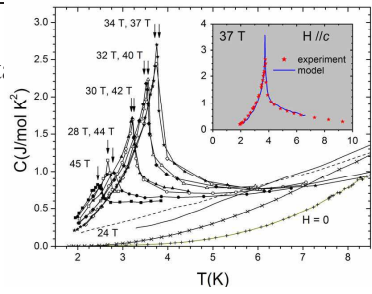}
\vspace{-0.8cm}
\caption{(color online)Specific heat vs temperature at constant magnetic
fields $H$. A low temperature $\protect\lambda$-anomaly is evident at $H
\geq 28~\mathrm{T}$. The anomaly first moves to higher temperatures with
increasing magnetic fields, it reaches a maximum at $H = 36 \pm 1~\mathrm{T}$%
, and then decreases for $37 < H < 45~\mathrm{T}$. Inset: Specific heat vs $T$ at $H = 37 \mathrm{%
T}$ after subtraction of a small contribution due to the $S^z = 0$ triplet
level only relevant at higher temperatures, and phonons. Also displayed is
the result of our Monte Carlo calculation.}
\label{fig3}
\end{figure}

The observation of a phase transition is further confirmed by specific heat
measurements as a function of temperature for constant magnetic fields shown
in Figure~3. This experiment was performed in a calorimeter made
out of plastic materials \cite{Jaime00} using both a superconducting magnet
and the $45~\mathrm{T}$ hybrid magnet at the National High Magnetic Field
Laboratory. At zero magnetic field the specific heat is featureless. As soon
as finite external field along the $c$-axis is applied the specific heat
increases due to the reduction in the singlet-triplet energy gap $\Delta (H)$%
. A small anomaly then develops for $H=28~\mathrm{T}$ at $T\sim 2.7~\mathrm{K%
}$ that moves to higher temperatures as the magnetic field is further
increased. This $\lambda $-shaped anomaly also grows with magnetic field due
to the increased number of triplons (magnetization). Their number at low
temperatures is roughly proportional to the magnetic field (see Figure~2(a)). 
Once the middle of the magnetization ramp is reached at $H=36\pm 1~%
\mathrm{T}$, the transition temperature and the size of the $\lambda $%
-anomaly then start to decrease owing to the reduction in the availability
of singlet dimers. Moreover, the mirror symmetry of the phase diagram around
this maximal point (see Fig. 4) offers an experimental confirmation of the 
\emph{particle-hole symmetry} implicit in $H_{\text{\emph{eff}}}$.
No phonon contribution has been subtracted from the specific heat data. The
inset of figure~3 displays the data at $37~\mathrm{T}$ after
subtraction of a small exponential contribution (activated energy $\Delta
(H=0)=3.13$~meV) from the $S^{z}=0$ triplet components that becomes apparent
only above 8~K. In addition, a Debye phonon contribution with $\Theta
_{D}=350~\mathrm{K}$ was subtracted. The solid line is the result of our
quantum Monte Carlo calculation for $H_{\text{\emph{eff}}}$ after a finite
size scaling to the thermodynamic limit. The systems sizes that were used
for the scaling are: $L=4,6,8,12$.

\begin{figure}[htb]
\vspace{-2.2cm} 
\includegraphics[angle=270,width=6.5cm,scale=1]{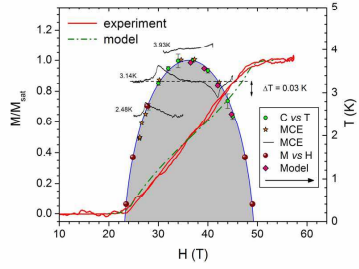}
\vspace{-0.8cm}
\caption{(color online) Left $y$-axis: Magnetization normalized to the
saturation value (M/M$_{sat}$) vs magnetic field along the c-axis, measured
at $1.5~\mathrm{K}$ (red line). Results for the model discussed in the text
(green line). Right $y$-axis: Transition temperature from specific heat vs
temperature, magnetocaloric effect (MCE) and magnetization vs field data.
Black lines are the sample temperature measured while sweeping the magnetic
field quasi-adiabatically.}
\label{fig4}
\end{figure}

The phase diagram of figure~4 shows all our experimental data
combined, together with some temperature traces measured while sweeping the
magnetic field at a rate of about $12 \mathrm{T}/\mathrm{min}$, i.e.
magneto-caloric effect (MCE) \cite{Jaime02}. The left side $y$-axis is for
the magnetization versus $H$ curve. From this curve we obtain: $H_{c1}= 23.5~%
\mathrm{T}$ and $H_{c2} = 49~\mathrm{T}$. The right side $y$-axis is for the
transition temperature versus field plot obtained from the specific curves
(see figure~3) and the magneto-caloric effect measurements. The
anomalies in the MCE, observed upon crossing the phase boundaries in the
direction of increasing fields, evidence reduced magnetic entropy within the
critical region consistent with a gas-to-liquid phase transition. Also
displayed in figure~4 are the transition points calculated with our
Monte Carlo algorithm for $L=12$. The BEC phase is represented by the shaded
region whose boundary is intended merely as a guide to the eye. The
excellent agreement between experiment and theory is a compelling case for
BaCuSi$_2$O$_6$ being a true realization of a quasi two dimensional BEC of
magnetic degrees of freedom.

Early Chinese chemists could not have imagined two thousand years ago that
BaCuSi$_{2}$O$_{6}$ was not only an attractive purple pigment but also a
potential solid state device for exploring the quantum effects of a BEC at
liquid $^{4}$He temperatures in magnetic fields. Indeed, we show that a
minimal Hamiltonian for a lattice gas of hard core bosons describes
surprisingly well the magnetization and specific heat of this magnetic
system in external magnetic fields. To derive an effective Hamiltonian we
simply neglected the high energy triplet states $|\phi _{\mathbf{i}%
}^{t},S^{z}=0\rangle $ and $|\phi _{\mathbf{i}}^{t},S^{z}=-1\rangle $. The
excellent agreement with the experimental data, and the verification of the
particle-hole symmetry implicit in $H_{\text{\emph{eff}}}$ validate our
approach. However, other physical properties such as the small staggered
magnetization that appears in the \emph{ab} plane, when the triplets
condense, would require the inclusion of virtual processes that involve the
above mentioned states. It is also important to remark that the phase
coherence of the spin BEC requires the conservation of the number of
particles that is equivalent to the total magnetization. This condition is
fulfilled whenever the crystal has some degree of rotational symmetry about
the direction of the applied field like, for instance, a four-fold symmetry axis.
Otherwise, the presence of anisotropy terms will significantly reduce the
de-coherence time of the condensate. Finally, the weakly coupled Cu-bilayers
in the crystal structure of BaCuSi$_{2}$O$_{6}$ open the door for future
studies like the dimensional crossover of a bosonic gas. By
increasing the distance between adjacent bilayers with chemical
substitutions the $\lambda $-shaped second order phase transition,
characteristic of a three-dimensional system, is replaced by a
Kosterlitz-Thouless phase transition.

We thank E.W. Fitzhugh for references to early work on Ba-Cu silicates, P.
Littlewood, D. Pines and A. Abanov for helpful discussions. Experiments
performed at the National High Magnetic Field Laboratory were supported by
the U.S. National Science Foundation through Cooperative Grant No.
DMR9016241, the State of Florida, and the U.S. Department of Energy.

\end{document}